\documentclass[conference]{IEEEtran}

\usepackage{graphicx}
\usepackage{subcaption}
\usepackage{balance}
\usepackage{comment}
\usepackage{amsmath}
\usepackage{hyperref}
\usepackage{courier}
\usepackage[ruled,linesnumbered]{algorithm2e}

\usepackage{balance}

\usepackage{multirow}

\IEEEoverridecommandlockouts
\begin{document}

\title{A Study of Runtime Adaptive Prefetching for STTRAM L1 Caches}

	\author{\IEEEauthorblockN{Kyle Kuan and
	Tosiron Adegbija}
	\IEEEauthorblockA{Department of Electrical \& Computer Engineering \\ University of Arizona, Tucson, AZ, USA \\
	Email: \{ckkuan, tosiron\}@email.arizona.edu}
	\vspace{-15pt}
}

	
\maketitle

\begin{abstract}
Spin-Transfer Torque RAM (STTRAM) is a promising alternative to SRAM in on-chip caches due to several advantages. These advantages include non-volatility, low leakage, high integration density, and CMOS compatibility. Prior studies have shown that relaxing and adapting the STTRAM retention time to runtime application needs can substantially reduce overall cache energy without significant latency overheads, due to the lower STTRAM write energy and latency in shorter retention times. In this paper, as a first step towards efficient prefetching across the STTRAM cache hierarchy, we study prefetching in reduced retention STTRAM L1 caches. Using SPEC CPU 2017 benchmarks, we analyze the energy and latency impact of different prefetch distances in different STTRAM cache retention times for different applications. We show that \textit{expired\_unused\_prefetches}---the number of unused prefetches expired by the reduced retention time STTRAM cache---can accurately determine the best retention time for energy consumption and access latency. This new metric can also provide insights into the best prefetch distance for memory bandwidth consumption and prefetch accuracy. Based on our analysis and insights, we propose \textit{Prefetch-Aware Retention time Tuning (PART)} and \textit{Retention time-based Prefetch Control (RPC)}. Compared to a base STTRAM cache, PART and RPC collectively reduced the average cache energy and latency by 22.24\% and 24.59\%, respectively. When the base architecture was augmented with the state-of-the-art near-side prefetch throttling (NST), PART+RPC reduced the average cache energy and latency by 3.50\% and 3.59\%, respectively, and reduced the hardware overhead by 54.55\%.

\end{abstract}

\section{Introduction}
Much research has focused on optimizing caches' performance and energy efficiency due to the caches' non-trivial impact on processor architectures. These optimization efforts are especially important for resource-constrained devices for which low-overhead energy reduction remains a major concern. An increasingly popular approach for improving caches' energy efficiency involves replacing the traditional SRAM with emerging non-volatile memory (NVM) technologies. 

Among several NVM alternatives, Spin-Transfer Torque RAM (STTRAM) has emerged as a promising candidate for replacing traditional SRAMs in future on-chip caches. STTRAMs offer several attractive characteristics, such as non-volatility, low leakage, high integration density, and CMOS compatibility. However, some of STTRAM's most important challenges include its long write latency and high write energy \cite{DASCA,sayed2018cross}. These challenges are attributed, in part, to the STTRAM's long \textit{retention time}---the duration for which data is maintained in the memory in the absence of power. For caches, the intrinsic STTRAM retention time of up to 10 years is unnecessary, since most cache blocks need to be retained in the cache for no longer than 1\textit{s} \cite{Jog12}. Furthermore, different applications or application phases may have different retention time requirements \cite{kuan19energy}. Thus, prior research has proposed reduced retention STTRAMs that can be specialized to the needs of various applications \cite{kuan19energy} or different cache levels \cite{Sun11}. 

To further improve cache efficiency, cache prefetching is a popular technique that fetches data blocks from lower memory levels before the data is actually needed. While prefetching can be very effective for improving cache access time, inaccurate prefetching can cause cache pollution, increase memory bandwidth contention, and in effect, degrade the cache's performance and energy efficiency \cite{Srinath07,Heirman18}. Apart from determining the right prefetch targets, the prefetch distance must also be well-monitored such that it maintains good prefetch accuracy \cite{Heirman18}. This is especially important in reduced retention STTRAMs, which, our analysis show, exhibit different locality behaviors than traditional SRAM caches due to cache block expiration.

In this paper, as an important first step towards understanding prefetching across the STTRAM cache hierarchy, we study data prefetching in the context of a reduced retention L1 STTRAM cache---simply referred to hereafter as 'STTRAM cache'. We assume an STTRAM cache that features the ability to adapt to different applications' retention time requirements (e.g., \cite{Sun11,kuan19energy}). We focus on the potentials of data prefetching for improving STTRAM cache's energy efficiency. To motivate this study, we performed extensive experiments using a variety of SPEC 2017 benchmarks and a PC-based stride prefetcher that prefetches memory addresses based on the current program counter (PC) \cite{Chen95}. We observed that if earlier prefetched data blocks are expired because of the reduced retention time, a conventional prefetcher would not reload these blocks. However, a prefetcher could be modified to reload these blocks, thereby reducing the miss penalty caused by premature expiration of blocks (i.e., \textit{expiration misses} \cite{dhruv19}). Furthermore, the low write energy in reduced retention STTRAM also mitigates the negative impact of writing blocks in addition to demand requests. We also observed that common metrics for determining the best retention time during runtime (e.g., cache miss rates \cite{kuan19energy}) may not be accurate in the presence of a prefetcher and can unnecessarily waste energy. As such, prefetching, if carefully designed in the context of reduced retention STTRAMs, can increase energy savings as compared to prior reduced retention STTRAM design techniques, without incurring significant latency overheads.

Based on the above observations, we propose a new metric, which we call \textit{expired\_unused\_prefetches}, to evaluate the quality of a current retention time and prefetch distance. The \textit{expired\_unused\_prefetches} represents the number of prefetched blocks that were not accessed by a demand request before expiry. Using this metric,  we developed \textit{Prefetch-Aware Retention time Tuning (PART)} and \textit{Retention time-based Prefetch Control (RPC)}. During a brief runtime profiling phase for each application, PART uses the ratio of \textit{expired\_unused\_prefetches} to total prefetches to determine if the current retention time suffices for the application. The retention time selected by PART indicates the average amount of time for which cache blocks used by an application reside in the cache. As such, if too many prefetches are expired without being used, it is likely that those prefetches were inaccurate. RPC uses this idea to map \textit{expired\_unused\_prefetches} to the prefetch distance.

Our major contributions are summarized as follows:
\begin{description} 
\item[$\bullet$] We study prefetching in STTRAM caches and propose a metric---\textit{expired\_unused\_prefetches}---that can be used to effectively determine both retention time and prefetch distance, without the need for any complex hardware overhead.
\item[$\bullet$] Using \textit{expired\_unused\_prefetches}, we proposed an algorithm to determine retention time and prefetch distance during runtime. 
\item[$\bullet$] Compared to a base state-of-the-art reduced retention time STTRAM cache, PART+RPC reduced the average energy and latency by up to 22.24\% and 24.59\%, respectively. Furthermore, when the base architecture was augmented with the state-of-the-art near-side prefetch throttling (NST) prefetching, our approach reduced the average energy and latency by 3.50\% and 3.59\%, respectively, and substantially reduced the hardware overhead by 54.55\%.
\end{description}

\section{Background and Related Work}
STTRAM's basic structure, comprising of magnetic tunnel junction (MTJ) cells, and characteristics have been detailed in prior work \cite{Smullen11}. Earlier works suggest the use of very short retention times (e.g., 26.5 $\mu$s \cite{Sun11}) with a DRAM-style refresh scheme for cache implementation \cite{Sun11,Jog12}. Recent works show that adapting a set of pre-determined retention times to applications' needs, specifically the cache block lifetimes, can further improve energy consumption \cite{kuan19energy,kuan19halls}. In this section, we present a brief overview of prior work on adaptable retention time STTRAM cache---the architecture on which we build the analysis presented herein---and an overview of prefetch distance control.

\subsection{Adaptable Retention Time STTRAM Caches}
Recent optimizations on STTRAM cache exploit the variable cache block needs of different applications for energy minimization. For example, Sun et al. \cite{Sun11} proposed a multi-retention time cache featuring various retention times enabled by various MTJ designs, wherein different applications could be run on the retention time that suits them best. More recently, Kuan et al. \cite{kuan19energy} analyzed the retention times of different applications and proposed a logically adaptable retention time (LARS) cache \cite{kuan19energy} that used multiple STTRAM units with different retention times. LARS involves a hardware structure that samples the application's characteristics during its very first run. Based on the applications' retention time requirements, each application is executed on the retention time unit that best satisfies their retention time needs. In this paper, we assume a similar multi-retention time architecture to LARS. For brevity, we direct readers to \cite{kuan19energy} for additional low-level details of the architecture, but omit those details herein.
\vspace{-5pt}

\subsection{Prefetch distance control}
Prefetch distance refers to how far into a demand miss stream that a prefetcher can prefetch \cite{Chen95}. Effective prefetching relies on accurate prefetch addresses and timely arrival of data blocks to hide the latency between processor and main memory. As such, the prefetch distance must not be so short as to generate excessive \textit{late\_prefetches} \cite{Srinath07} or too long to lose prefetch accuracy \cite{Srinath07,Heirman18}. Inaccurate prefetches can cause performance degradation due to the saturation of memory bandwidth and cache pollution. As such, lots of prior works discuss various techniques for controlling prefetch distance, feedback directed prefetching techniques, ways to monitor the number of total prefetches and late prefetches to evaluate prefetch accuracy and lateness, and how to determine the prefetcher aggressiveness. For example, Ebrahimi et al. \cite{Ebrahimi09Coor} proposed a rule-based control method to separate global throttling and local throttling, and  reduce inter-core interference. Both \cite{Ebrahimi09Coor} and \cite{Srinath07} looked at the number of useless prefetches, which is determined by prefetches that are not used before they are evicted. Heirman et al. \cite{Heirman18} referred to the aforementioned methods as \textit{far\-side} throttling, since they maintained high prefetch distance and throttled down when negative effects were observed. Heirman et al. \cite{Heirman18} proposed near-side prefetch throttling (NST), which monitored the ratio of late prefetches and total prefetches, kept prefetch distance low and only raised the distance if necessary. None of these techniques, however, have considered prefetching in STTRAM caches. As we show in our analysis herein, state-of-the-art prefetchers may under-perform if simply implemented on STTRAM caches without considering execution characteristics and metrics that are unique to STTRAM caches.

\begin{figure}[h]

\begin{subfigure}{.48\linewidth}

		\centering
		\includegraphics[width=\linewidth]{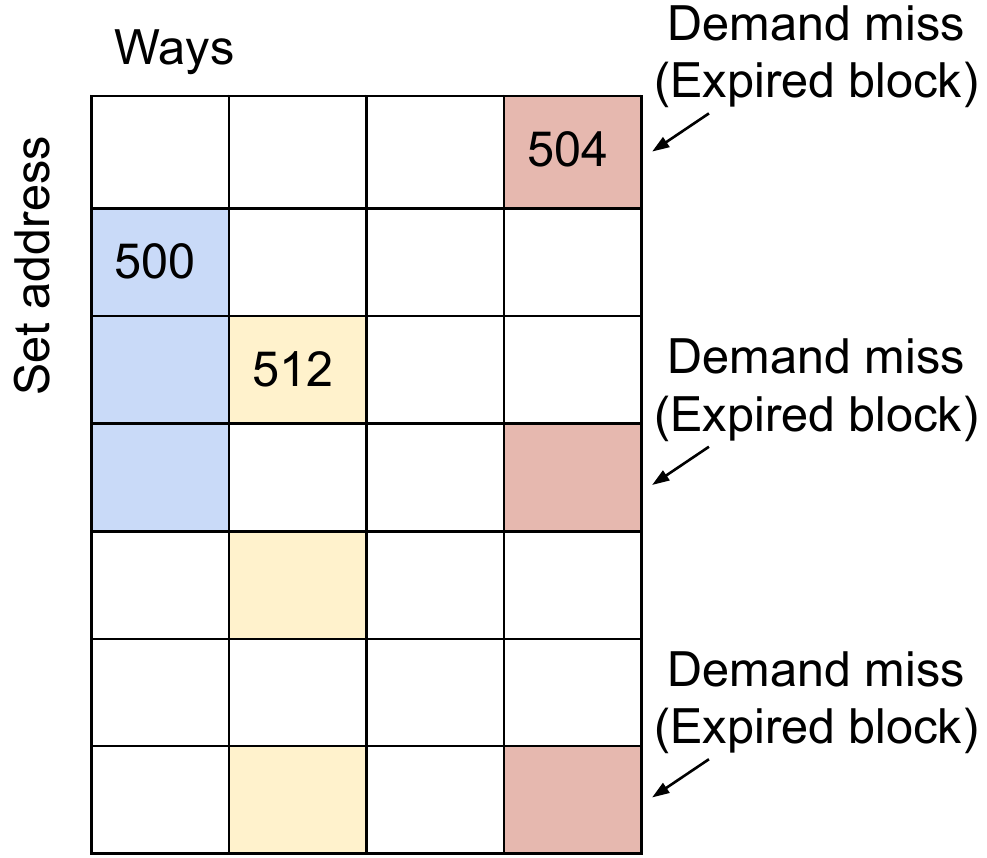}
		\caption{Prefetch disabled}
		\label{fig:npfexp}
\end{subfigure}
~
\begin{subfigure}{.48\linewidth}
		\centering
		\includegraphics[width=\linewidth]{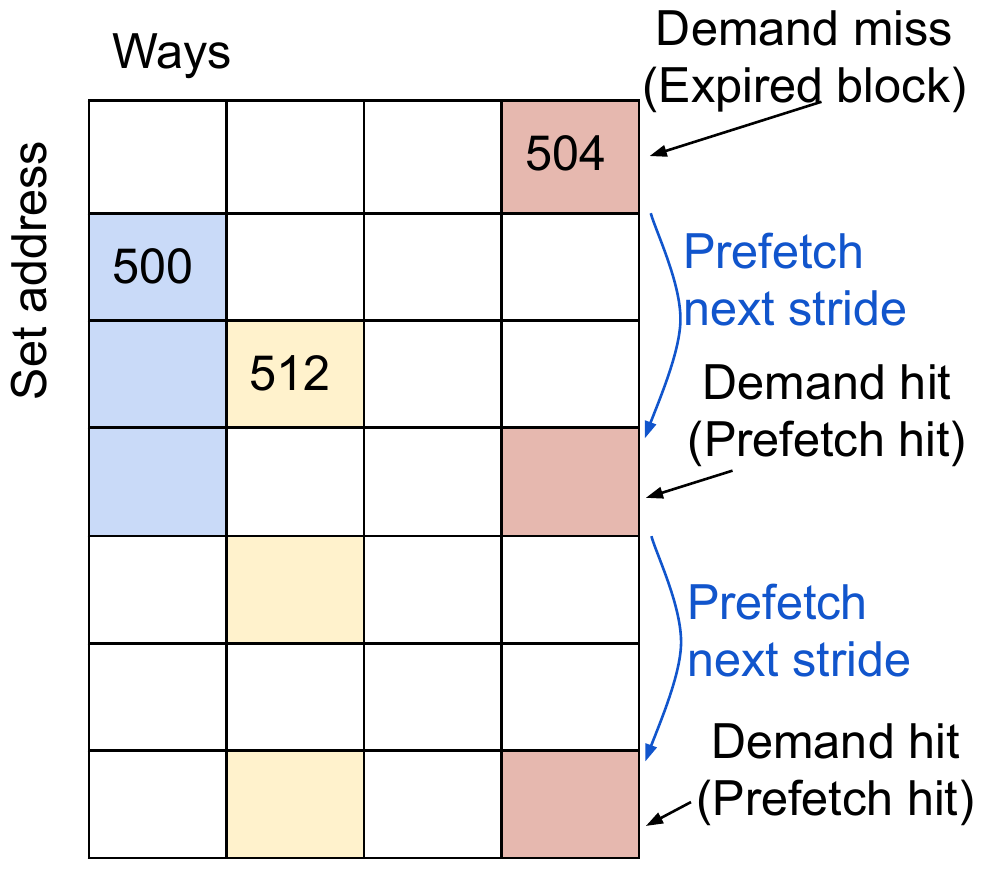}
		\caption{Prefetch enabled}
		\label{fig:pfexp}
\end{subfigure}
\caption{Prefetching expired blocks. In (a) the prefetch does not bring back previously expired blocks into the cache; in (b) the previously expired blocks are brought back into the cache}
\label{fig:prefetch_exp}
\end{figure}

\section{Enabling Prefetching in STTRAM Cache}

\subsection{Effectiveness of prefetching expired blocks} \label{sec:npfexp}
Expired blocks in STTRAM caches incur misses when a demand request accesses an expired block prior to eviction. We refer to these misses as \textit{expiration misses}, similar to prior work \cite{dhruv19}. As the retention time becomes shorter, expiration misses increase, until expiration misses become the majority of misses and essentially disables the cache's ability to exploit temporal locality. Given the uniqueness of expiration misses in STTRAM caches, we first studied the impact of prefetching on expired cache blocks. Figure \ref{fig:prefetch_exp} illustrates a simplified diagram of a data cache, with each cell representing a cache block. The horizontal blocks represent the cache ways (four ways in total) and the vertical blocks represent the set address (seven set addresses in total).  The blocks' colors represent the prefetch stream that brought the cache blocks into the cache. We used the stride prefetcher \cite{Chen95} as the base to illustrate our idea and in our experiments. The number associated with the color represents the program counter (PC) value of the load/store instruction that begins the stream due to a demand miss. 

Figure \ref{fig:npfexp} illustrates the STTRAM cache without prefetching expired blocks. Assume that the instruction at PC 504 brought three cache blocks into the cache. Since the blocks are brought in by the same stream, they are likely to expire around the same time. If the prefetcher is disabled on those expired blocks, as in a conventional prefetcher, when the demand request accesses the blocks again, loading each block will incur the miss penalty due to expiration misses. Alternatively, enabling the prefetcher for the expired blocks can have a positive effect, since, as shown in Figure \ref{fig:pfexp}, the prefetcher brings in subsequent blocks after the first demand miss (expiration miss). Thus, subsequent accesses to the prefetched blocks become demand hits without exposing the memory latency.

\begin{figure}[ht]
        \centering
		\includegraphics[width=\linewidth]{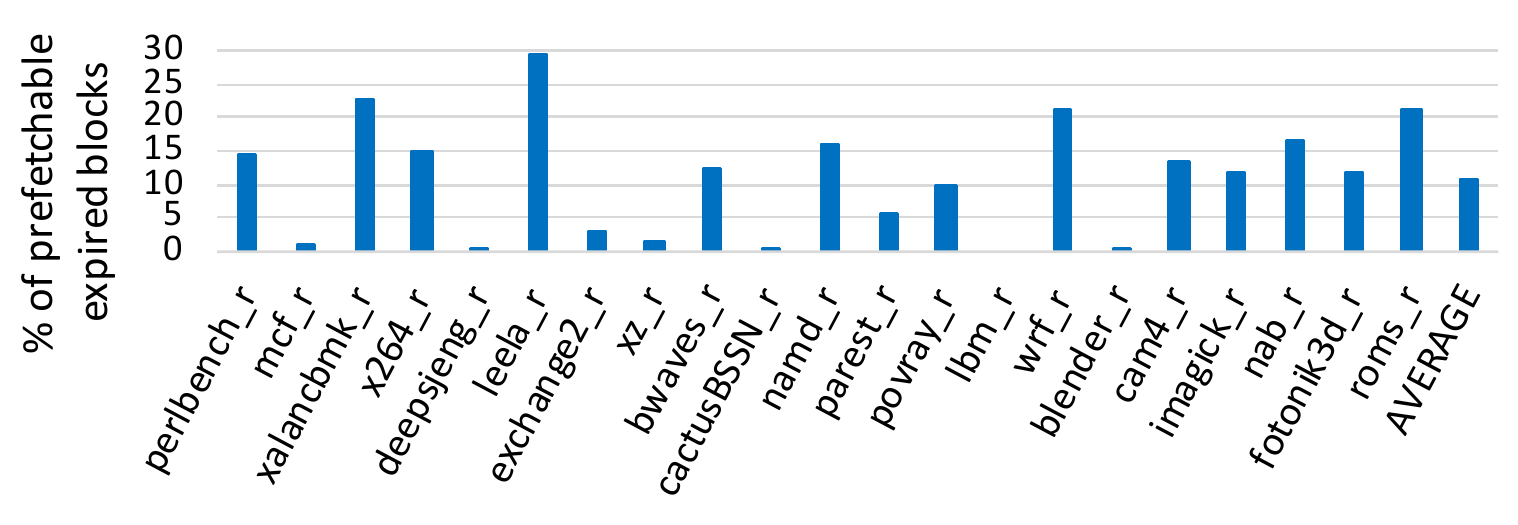}
        \caption{Percentage of prefetchable expired blocks in total expired blocks across SPEC CPU 2017 benchmarks}
\label{fig:pfexp_total}
\end{figure}

\begin{figure}[ht]
        \centering
		\includegraphics[width=0.65\linewidth]{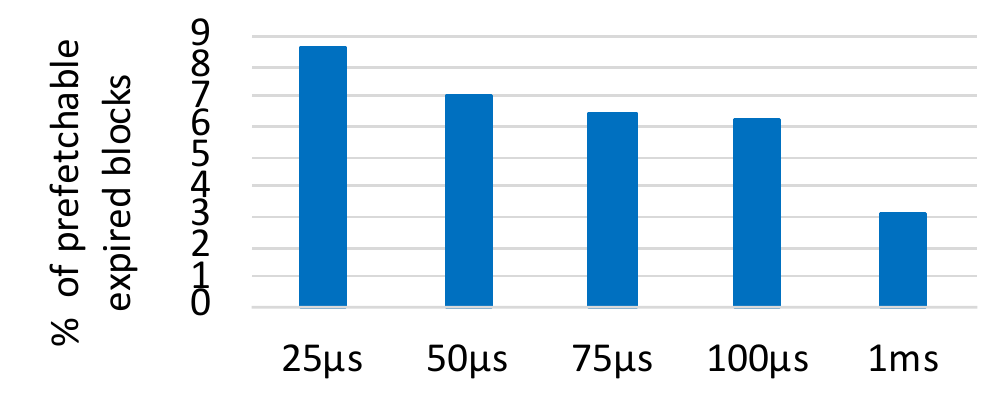}
		\vspace{-5pt}
\caption{Percentage of prefetchable expired blocks in total expired blocks for different retention times for SPEC CPU 2017 benchmarks (Geometric mean is shown for brevity)}
\label{fig:pfexp_total_retention}
\vspace{-10pt}
\end{figure}

To quantify the benefits of prefetching expired blocks, we performed experiments using SPEC CPU 2017 \textit{rate (\_r)} benchmarks and evaluated the energy and latency changes. We used a base stride prefetcher of prefetch distance 16, similar to \cite{arm_cortexa72_technical} and considered retention times from 25$\mu$s to 1ms. Our detailed simulation setup is described in Section \ref{sec:setup}. We use the term \textit{prefetchable expired blocks} to represent expired blocks that can be accurately predicted and reloaded through the stride prefetcher, and therefore would incur no expiration miss. Figure \ref{fig:pfexp_total} shows the percentage of prefetchable expired blocks in total expired blocks across the benchmarks, assuming the best retention times. On average across all benchmarks, 10.85\% of expired blocks can be reloaded into the cache for reuse. Depending on the applications' access pattern and cache block lifetimes, the reused expired blocks can be as high as 29.66\% for \textit{leela}, while over the half of benchmarks (13 of 21) have reuse rates over 10\%. To further illustrate this behavior, Figure \ref{fig:pfexp_total_retention} shows the percentage of prefetchable expired blocks in total expired blocks for different retention times. For brevity, the geometric mean is shown for the different retention times. In general, the percentage of reused expired blocks increases as the retention time decreases, with the highest being 8.69\% at 25 $\mu$s. These analysis motivate us to explore low-overhead techniques for prefetching \textit{and} determining the best retention time in STTRAM caches during runtime.

\begin{figure}[ht]
		\vspace{-15pt}
		\centering
		\includegraphics[width=\linewidth]{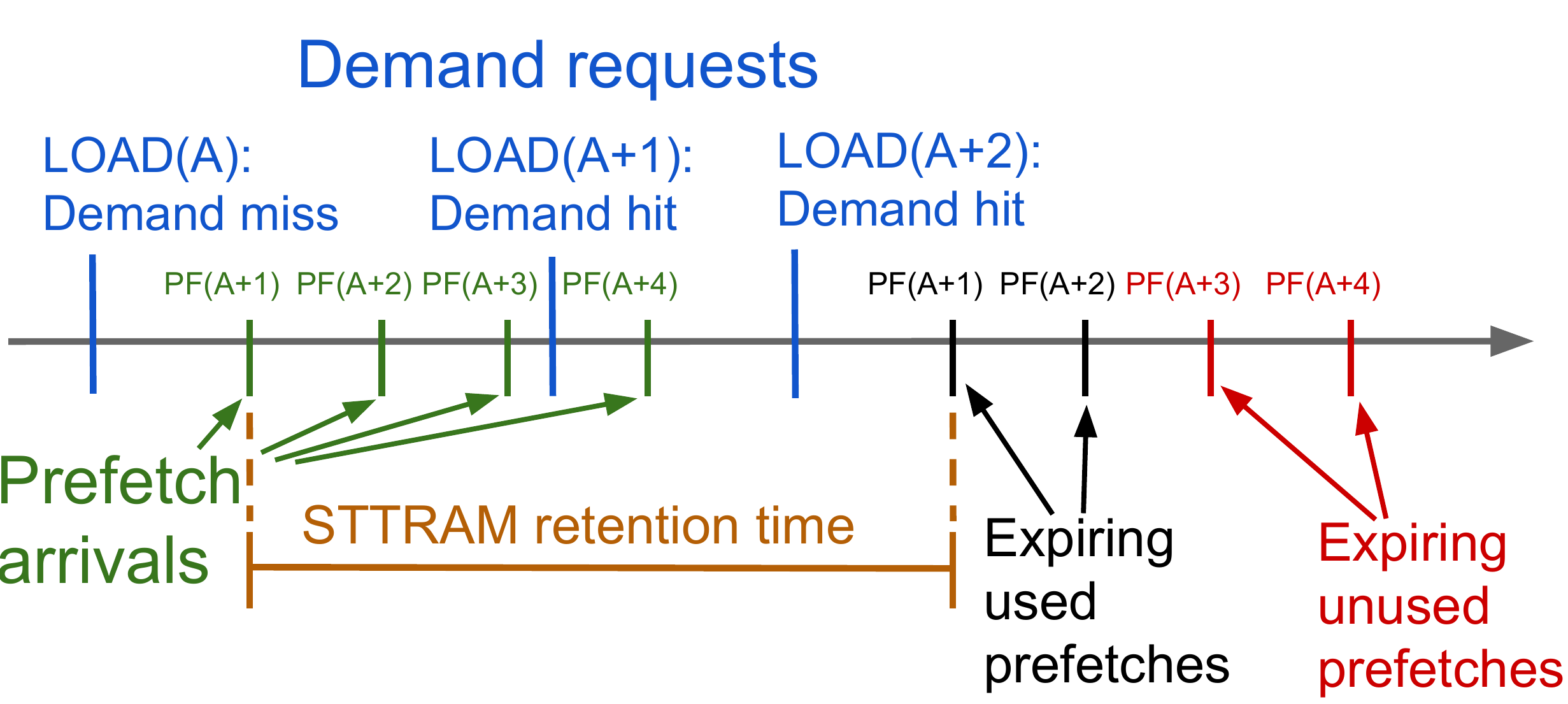}
		\caption{Retention time expiration detect potentially unused prefetches}
		\label{fig:expired_unused}
        \vspace{-15pt}
\end{figure}

\begin{algorithm}[t]
\SetDataSty{small}
\SetKwData{expiredPF}{expiredPF}
\SetKwData{baseExpiredPF}{baseExpiredPF}
\SetKwData{allPF}{allPF}
\SetKwData{CurMissRate}{CurMissRate}
\SetKwData{OutputRetentionTime}{OutputRetentionTime}
\SetKwFunction{expiredUnusedPrefetches}{expiredUnusedPrefetches}
\SetKwFunction{totalPrefetches}{totalPrefetches}
\SetKwFunction{totalMSHRRequests}{totalMSHRRequests}
\SetKwFunction{missBasedTuning}{missBasedTuning}
\KwData{Retention time set  $R=\{25\mu s, 50\mu s, 75\mu s, 100\mu s,1ms\}$}
\KwResult{OutputRetentionTime}
\BlankLine

\OutputRetentionTime $\leftarrow 1ms$\;
\ForEach{$r \in R$}{
    \allPF$\leftarrow$\totalPrefetches{$r$}$/$
    \totalMSHRRequests{$r$}\;\label{line:allPF}
    \expiredPF $\leftarrow$ \expiredUnusedPrefetches{$r$}$/$
    \totalPrefetches{$r$}\;\label{line:expiredPF}
    \If{\allPF $>$ 0.1\%}{
        \If{\baseExpiredPF is set}{
            \If{\expiredPF $<$ 2*\baseExpiredPF}{\label{line:tuningBegin}
            \OutputRetentionTime $\leftarrow r$\;
            }
            \Else{
                \Return{\OutputRetentionTime}\;
            }\label{line:tuningEnd}
        }
        \Else{ 
            \OutputRetentionTime $\leftarrow r$\;  \label{line:setbaseBegin}
            \If{\expiredPF $>$ 0.02\%}{
            \baseExpiredPF $\leftarrow$ \expiredPF\;
            }\label{line:setbaseEnd}
        }
    }
    \Else{
        \OutputRetentionTime $\leftarrow r$\;
        \missBasedTuning{\OutputRetentionTime}\; \label{line:MissCond}
        \Return{\OutputRetentionTime}\;
    }
}
\Return{\OutputRetentionTime}\;

\caption{Prefetch-Aware Retention Time Tuning}
\label{algo:algo1}
\end{algorithm}

\subsection{Prefetch-Aware Retention time Tuning (PART)} \label{sec:PART}
A key point of our analysis so far is that, as illustrated in Figure \ref{fig:pfexp}, expiration of cache blocks must be considered in the design of prefetchers. Furthermore, we also analyzed prior adaptable retention time techniques (e.g., \cite{kuan19energy}) that used miss rates to predict the best retention time. We found that these techniques only accurately predicted the best retention time using cache miss rates in the absence of a prefetcher. When a prefetcher is introduced, using miss rates may not be as accurate due to the interplay of expiration misses and prefetching. Thus, we designed the \textit{prefetch-aware retention time tuning (PART)} technique to take into account the expiration misses.

To motivate PART, Figure \ref{fig:expired_unused} illustrates the timeline of when prefetched blocks are brought into the cache and then expired. Assume that \texttt{LOAD(A)} instruction accesses memory address \textit{A} and causes a demand miss, the prefetcher sends out four requests from address \textit{A+1} to \textit{A+4}. The prefetch arrival times are marked in green color. After the retention time elapses, prefetched blocks begin to expire. We record the number of blocks that were not used by demand requests before expiration; we refer to these blocks as \textit{expired\_unused\_prefetches}. The basic idea of PART is to use the shortest retention time that does not excessively increase the \textit{expired\_unused\_prefetches}. To this end, PART tracks the changes in \textit{expired\_unused\_prefetches} at prefetch degree 1 during different tuning intervals to determine the best retention time. 

Algorithm \ref{algo:algo1} depicts the PART algorithm, which takes as input the available retention times in the system and outputs the best retention time. PART iterates through the available retention time set starting from the longest to the shortest (e.g., 1ms to 25$\mu$s), runs the application for a sampling period, and takes the ratio of total prefetches to total MSHR requests (\textit{allPF}) and the ratio of \textit{expired\_unused\_prefetches} to total prefetches (\textit{expiredPF}), as shown in lines \ref{line:allPF}-\ref{line:expiredPF}. If \textit{allPF} is smaller than 0.1\%, we infer that prefetches do not substantially contribute to memory traffic. Therefore, the algorithm switches to a subroutine that predicts the retention time based on cache misses, similar to prior techniques \cite{kuan19energy} (line \ref{line:MissCond}). If \textit{allPF} is greater than 0.1\%, the algorithm first checks if \textit{expiredPF} is significant enough ($>$ 0.02\%). If \textit{expiredPF} is greater than 0.02 \%, this \textit{expiredPF} is stored as \textit{baseExpiredPF} and used in subsequent tuning stages. Otherwise, PART iterates the next available retention times to see if the thresholds are satisfied (line \ref{line:setbaseBegin}-\ref{line:setbaseEnd}). Note that we determined the thresholds empirically through extensive experiments and analysis. After obtaining \textit{baseExpiredPF}, PART explores shorter retention times to find the one that does not excessively increase \textit{expiredPF} as compared to \textit{baseExpiredPF}. PART checks if \textit{expiredPF} is smaller than twice \textit{baseExpiredPF}. If so, it proceeds to the next shorter retention time, otherwise, the current retention time is returned as the tuning result (line \ref{line:tuningBegin}-\ref{line:tuningEnd}).

\subsection{Retention Time-based Prefetch Control (RPC)} 
We also developed a simple heuristic, called \textit{retention time-based prefetch control (RPC)}, that works in conjunction with PART to determine the best prefetch distance during runtime. To minimize tuning overhead, RPC determines the best prefetch distance in 'one-shot' along with the retention time tuning by the PART algorithm. PART tracks \textit{expired\_unused\_prefetches} at prefetch degree 1 for tuning the retention time. The determined retention time represents the period that suffices, on average, for the executing applications' cache block lifetimes. A prefetch degree of 1 is usually considered conservative in prefetch distance throttling \cite{Srinath07, Ebrahimi09Coor}. As such, if \textit{expired\_unused\_prefetches} is excessively high after retention time tuning, it is likely that wrong addresses were prefetched. In this case, we maintain the prefetch distance of 1 to minimize cache pollution and memory bandwidth contention. RPC takes \textit{expiredPF} in Algorithm \ref{algo:algo1} as input to determine the prefetch aggressiveness, and maps the prefetch distance similarly to \cite{Srinath07}. Table \ref{tab:RPC} shows the distribution of this mapping, representing different ranges of \textit{expiredPF} and the associated prefetch distance. If \textit{expiredPF} is above 5\%, the stride pattern does not match the current application's data access. Thus, the prefetch distance is kept at 1 in order to maintain prefetch functionality. On the other extreme, we observed that some applications have the lowest \textit{expiredPF} (and energy consumption) at prefetch distance 32, which indicates that the stride prefetcher captures the applications' data access pattern and is able to recover expired blocks. 

\begin{table}[ht]
\vspace{-5pt}
\small  
\renewcommand{\arraystretch}{0.95}
\caption{Prefetch distances for different \textit{ExpiredPF}}
\label{tab:RPC}
\centering
\begin{tabular}{|c|c|}
    \hline
    \textit{ExpiredPF} at prefetch degree 1	& Prefetch distance\\
    \hline
    \hline
    Above 5\%	&1\\
    \hline
    1.01\% - 5\%&4\\
    \hline
    0.51\% - 1\% &8\\
    \hline
    0.05\% - 0.5\% &16 \\
    \hline
    Below 0.05\% &32 \\
    \hline
\end{tabular}
\end{table}

\begin{table*}[ht]
 \small  
\renewcommand{\arraystretch}{0.8}
\caption{Prefetcher configuration and STTRAM cache parameters with different retention times}
\vspace{-5pt}
\label{tab:parameters}
\centering
\begin{tabular}{c||cccccc}
    \hline
    Prefetcher Configuration		&\multicolumn{6}{c}{Type: stride prefetcher, degree: 4, adaptable prefetch distance: 1, 4, 8, 16, 32}\\
    \hline
    \hline
    Cache Configuration				&\multicolumn{6}{c}{ 32KB, 64B line size, 4-way, 22nm technology}\\
    \hline
    \hline
    Memory device				&SRAM       & STTRAM-25$\mu$s  &STTRAM-50$\mu$s	    &STTRAM-75$\mu$s    &STTRAM-100$\mu$s   &STTRAM-1ms\\
    \hline
    Write energy (per access) 	&0.002nJ    &0.006nJ	        &0.007nJ        &0.007nJ        &0.008nJ        &0.011nJ\\
    \hline
    Hit energy (per access)	    &0.008nJ    &\multicolumn{5}{c}{0.005nJ}\\
    \hline
    Leakage power               &75.968mW   &11.778mW           &11.778mW       &11.778mW       &11.778mW       &11.365mW\\
    \hline
    Hit latency (cycles)    	&2          &\multicolumn{5}{c}{1}\\
    \hline
    Write latency (cycles)    	&2          &2			        &3		        &3		        &3                  &4\\
    \hline
\end{tabular}
\end{table*}

\subsection{Overhead} \label{sec:overhead}
Assuming a base architecture that has the capability of multiple retention times (e.g., \cite{kuan19energy}), PART's major advantage is that it imposes negligible hardware and tuning overhead. PART exploits most of the hardware components described in \cite{kuan19energy} for tuning. In addition to the four 32-bit registers and one division circuit used in prior work, PART only requires one additional 32-bit register for \textit{allPF} and \textit{expiredPF}. To keep track of expired unused prefetches, PART only requires one custom hardware counter, which increments when an expiring block's prefetch bit is valid. Using the shared \textit{expiredPF} in PART, RPC requires only one 32-bit comparator. In total, we estimate that the area overhead is less than 1\% of modern processors like ARM Cortex-A72 \cite{arm_cortexa72_technical}.

We note that the base architecture incurs energy and latency switching overheads from migrating the cache state from one STTRAM unit to another. Switching occurs when an application is first executed during its sampling period. For example, given a tuning interval of 10 million instructions and five retention time options, sampling would require 50 million instructions. However, PART does not increase the switching overhead with respect to the base. In the worst case, each migration takes approximately 2560 cycles and 8.192nJ energy, resulting in total time and energy overheads of 10240 cycles and 32.768nJ, respectively. While these overheads are minimal in the context of full application execution, we reiterate that PART did not contribute to this overhead.

\section{Simulation Results}

\subsection{Experimental Setup} \label{sec:setup}

To perform our analysis and evaluate PART, we implemented PART using an in-house modified\footnote{The modified GEM5 version can be found at \url{www.ece.arizona.edu/tosiron/downloads.php}} version of the GEM5 simulator \cite{gem5}. We modified GEM5 \cite{gem5} to model cache block expiration, variable tag lookup and cache write latency, variable retention time units, and variable prefetch distance as described herein. To enable rigorous comparison of PART against the state-of-the-art, we used two recent prior works to represent the state-of-the-art---LARS \cite{kuan19energy} to represent adaptable retention time and NST \cite{Heirman18} to represent variable prefetch distance. We also implemented these two techniques in GEM5. We used configurations similar to the ARM Cortex A72 \cite{arm_cortexa72_technical}, featuring a 2GHz clock frequency, and a private L1 cache with separate instruction and data caches. For this work, we focused on data cache prefetching, since it provides much opportunity for runtime adaptability, as opposed to the instruction cache \cite{kuan19energy}. Every MSHR request from the L1 data cache is directly sent to an 8GB main memory, and incurs memory latency. We intend to explore the impact of our work on the instruction cache and lower level caches in future work. 

We considered five retention times: 25$\mu$s, 50$\mu$s, 75$\mu$s, 100$\mu$s, 1ms, which we empirically found to be sufficient for the considered benchmarks. We used the MTJ modeling techniques proposed in \cite{Chun13} to model the different retention times, and used NVSim \cite{NVSim} to estimate the energy for the different retention times. Table \ref{tab:parameters} depicts prefetcher configurations and the STTRAM cache parameters used in our experiments as obtained from the modeling tools and techniques. We used twenty-one SPECrate CPU2017 benchmarks \cite{spec2017}, cross-compiled for the ARMv8-A instruction set architecture. Each benchmark was run using the \textit{reference} input sets for 1B instructions after restoring checkpoints from 240B instructions.

\subsection{Results and Comparisons} \label{sec:results}
In this section, we compare the cache energy and access latency benefits of our work to prior work in various prefetch distance control scenarios. We denote uniform prefetch distance 1 to 32 as \textit{PFD\_N}, where N represents the memory address distance. RPC represents the optimal static distance among PFD\_N, since RPC accurately determines the distance in the sampling phase and uses that distance throughout the application's run. We use NST \cite{Heirman18} to represent the state-of-the-art dynamic prefetch distance throttling. 
We compare PART to the miss-based tuning algorithm used in LARS. We start with a direct comparison of PART to LARS without prefetching. Next, we compare PART to LARS with a uniform stride prefetcher and use moderate prefetch aggressiveness: prefetch degree 2 and prefetch distance 16 (LARS+PFD\_16), similar to prior work \cite{Srinath07}. Thereafter, we compare PART to LARS with the NST stride prefetcher (LARS+NST) to evaluate the improvement over dynamic prefetch distance throttling. Lastly, we compare PART to an SRAM cache with the NST stride prefetcher (SRAM+NST) to show the collective improvements of adaptable retention time STTRAM cache when prefetching is active. All energy and latency results of PART are normalized to the subject of comparison.

\subsubsection{Comparison to the base STTRAM cache (LARS)}

\begin{figure*}
    \begin{subfigure}{\linewidth}
		\centering
		\includegraphics[width=0.75\linewidth]{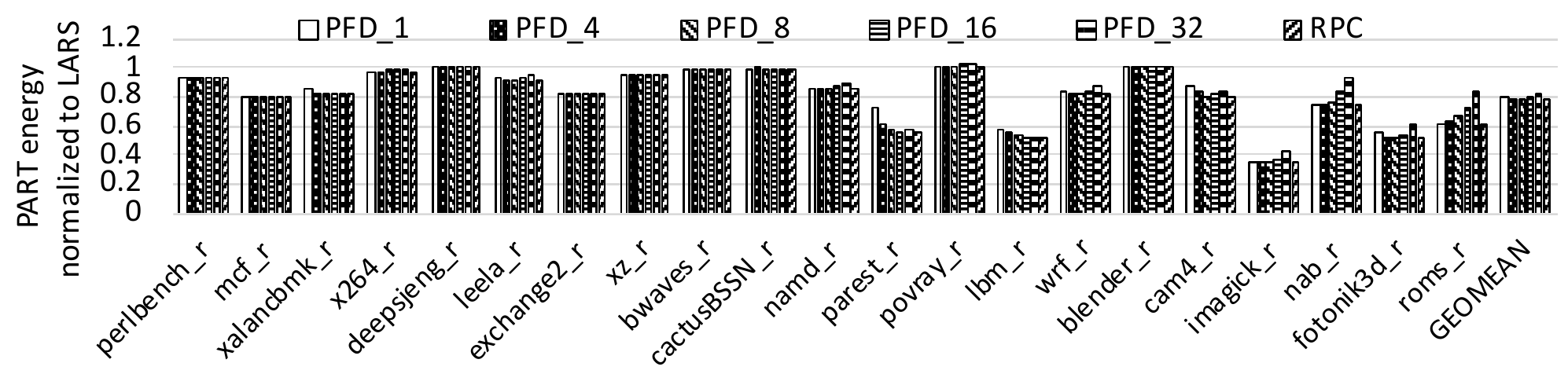}
		\caption{Energy}
		\label{fig:energy_lars}
    \end{subfigure}
    \begin{subfigure}{\linewidth}
		\centering
		\includegraphics[width=0.75\linewidth]{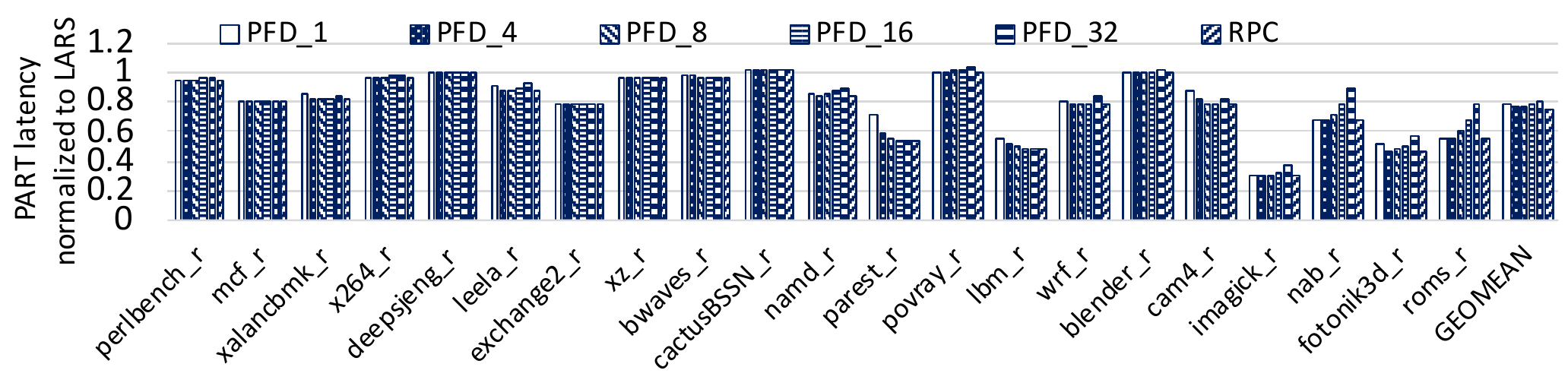}
		\caption{Latency}
		\label{fig:latency_lars}
    \end{subfigure}
    \vspace{-5pt}
    \caption{PART with different prefetch scenarios (PFD\_N and RPC) normalized to the base STTRAM cache (LARS)}   \label{fig:lars}
    \vspace{-10pt}
\end{figure*}
Figure \ref{fig:energy_lars} depicts the energy consumption of PART in different prefetch distance scenarios normalized to LARS. On average across all benchmarks, PART reduced the energy by 19.53\%, 21.25\%, 21.29\%, 20.09\%, and 17.68\% for PFD\_1, PFD\_4, PFD\_8, PFD\_16, and PFD\_32, respectively. RPC properly mapped expired unused prefetches (\textit{expiredPF}) to prefetch distance and ensured that the ideal static prefetch distance was selected. As such, PART+RPC reduced the average energy by 22.24\%, with savings as high as 65.96\% for \textit{imagick}. For \textit{parest}, \textit{imagick}, \textit{lbm}, \textit{roms}, and \textit{fotonik3d}, PART+RPC reduced the energy by more than 40\%, and no benchmarks' energy consumption was degraded by PART. Figure \ref{fig:latency_lars} depicts the cache access latency normalized to LARS without prefetching. On average across all benchmarks, PART reduced the latency by 21.52\%, 23.50\%, 23.51\%, 22.08\%,  19.29\%, and 24.59\% for PFD\_1, PFD\_4, PFD\_8, PFD\_16, PFD\_32, and RPC, respectively. PART+RPC reduced the latency by up to 70.41\% for \textit{imagick}. PART only incurred a negligible latency overhead (1.07\%) for \textit{cactusBSSN} while latency reductions were achieved for the rest of the twenty benchmarks.

Compared to LARS, we observed that the energy reduction trends were similar to the latency. Since prefetching can reduce compulsive misses, increased latency benefits are achieved as a result of the impact of expiration misses as discussed in Section \ref{sec:npfexp}. As shown in Figure \ref{fig:pfexp_total}, the average expired blocks that can be accurately prefetched and 'reused' are up to 10.85\%. Thus, the reduced expiration misses contributed significantly to miss latency reduction. 

\subsubsection{Comparison to LARS with uniform prefetch distance (LARS+PFD\_16)}
\begin{figure}[ht]
    \begin{subfigure}{\linewidth}
		\centering
		\includegraphics[width=0.9\linewidth]{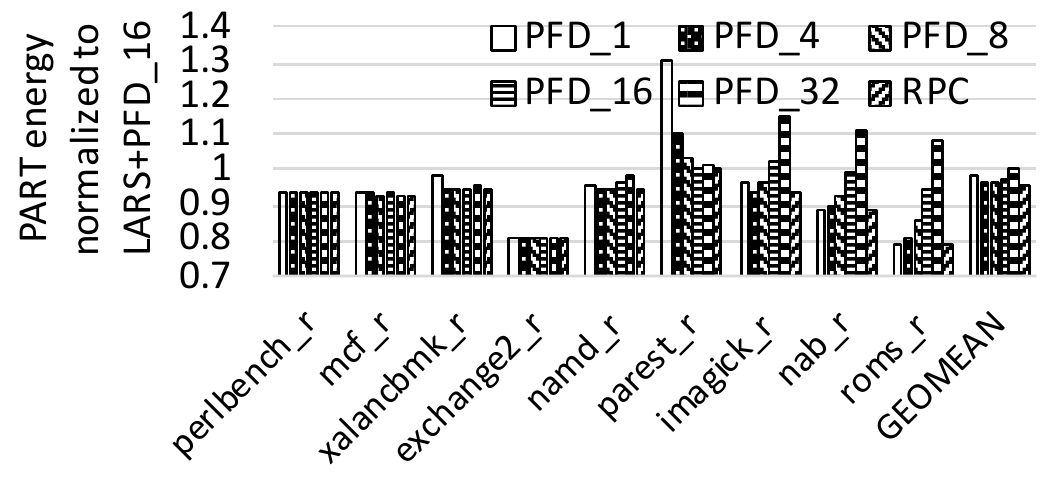}
		\caption{Energy}
		\label{fig:energy_lars16}
    \end{subfigure}
    \begin{subfigure}{\linewidth}
		\centering
		\includegraphics[width=0.9\linewidth]{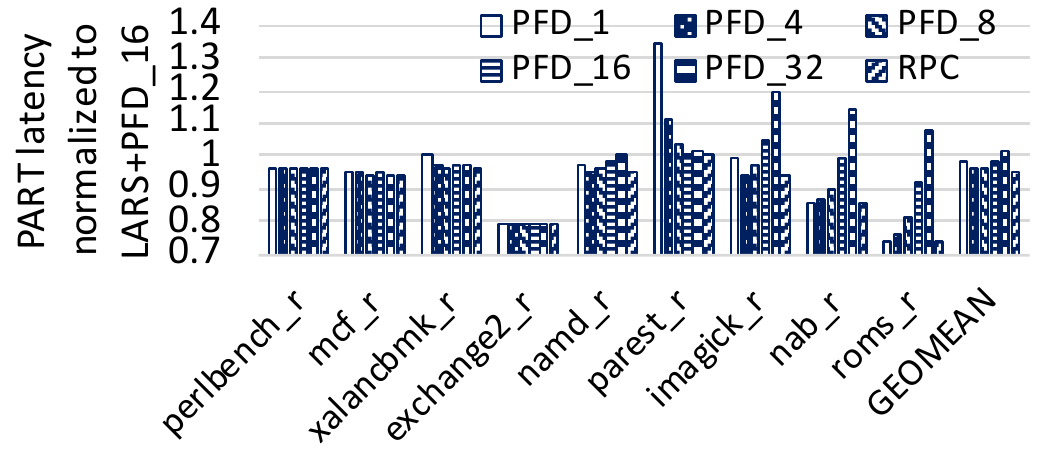}
		\caption{Latency}
		\label{fig:latency_lars16}
    \end{subfigure}
    \vspace{-5pt}
    \caption{PART with different prefetch scenarios (PFD\_N and RPC) normalized to LARS+PFD\_16}    \label{fig:lars16}
    \vspace{-15pt}
\end{figure}
Figure \ref{fig:lars16} depicts the energy and latency of PART normalized to LARS+PFD\_16. For brevity, only the geometric mean (across all the twenty-one benchmarks in Figure \ref{fig:lars}) and a subset of notable benchmarks are shown. Figure \ref{fig:energy_lars16} shows that across all the benchmarks, PART+RPC reduced the average energy consumption by 4.75\%, compared to LARS+PFD\_16 (the uniform prefetch distance). PART+RPC reduced the energy by up to 20.51\% and 18.77\% for \textit{roms} and \textit{exchange2}, respectively, with energy savings over 5\% for \textit{perlbench}, \textit{mcf}, \textit{xalancbmk}, \textit{namd}, \textit{nab}, and \textit{imagick}. We observed that PART generally selected shorter retention times than LARS+PFD\_16. By incorporating the expiration misses into the decision making about prefetching, PART achieved a balance of short retention times without translating into increases in miss latency. PART allowed the stride prefetcher to recover expired blocks in short retention times. PART+RPC only degraded the energy (by 0.48\%) for \textit{parest}. 

As described in Section \ref{sec:npfexp}, due to the reduced latency achieved by prefetching expired blocks, PART uses shorter retention times to improve energy consumption, since the short retention times do not substantially increase the latency. Figure \ref{fig:latency_lars16} shows that, similar to the energy improvement, PART+RPC reduced the average latency by 4.99\%, as compared to LARS+PFD\_16. PART+RPC reduced the latency by up to 25.76\%, 21.09\%, and 14.15\% for \textit{roms}, \textit{exchange2}, and \textit{nab}, respectively. To understand why PART performed so well for these benchmarks, we studied their execution more closely. For \textit{exchange2}, we observed that LARS selected a long retention time (1ms) due to low miss rates at 1ms, whereas shorter retention times increased the miss rates substantially (by up to 9x). However, the large amounts of misses at shorter retention times were rapidly amortized by stride prefetching and did not have substantial negative impact on the latency. We observed that even though shorter retention times increased \textit{totalPrefetches} for expired blocks, the \textit{expired\_unused\_prefetches} increased at a much slower rate, thereby substantially reducing \textit{expiredPF} (by up to 42\%). As such, PART selected short retention times (e.g., 25$\mu$s) and was able to improve the latency for these benchmarks. 

On the other hand, for \textit{roms}, LARS selected a short retention time of 25$\mu$s due to the low miss rates. However, the \textit{expiredPF} were substantially higher at shorter retention times than 1ms. As such, PART selected 1ms for \textit{roms} to save potentially useful prefetches with the longer retention time. The reduced latency in \textit{nab} resulted from the optimal prefetch distance (at PFD\_1) as determined by RPC. These results illustrate the importance of adaptable prefetch distance to satisfy different applications' needs. PART incurred minor latency overheads of up to 1.6\% and 0.19\% for \textit{cactusBSSN} and \textit{parest}, but reduced the latency for majority of the benchmarks (19 of 21).

\subsubsection{Comparison to LARS with dynamic prefetch distance (LARS+NST)}\label{sec:part_vs_larsnst}
\begin{figure}[b]
    \begin{subfigure}{\linewidth}
		\vspace{-10pt}
		\centering
		\includegraphics[width=0.9\linewidth]{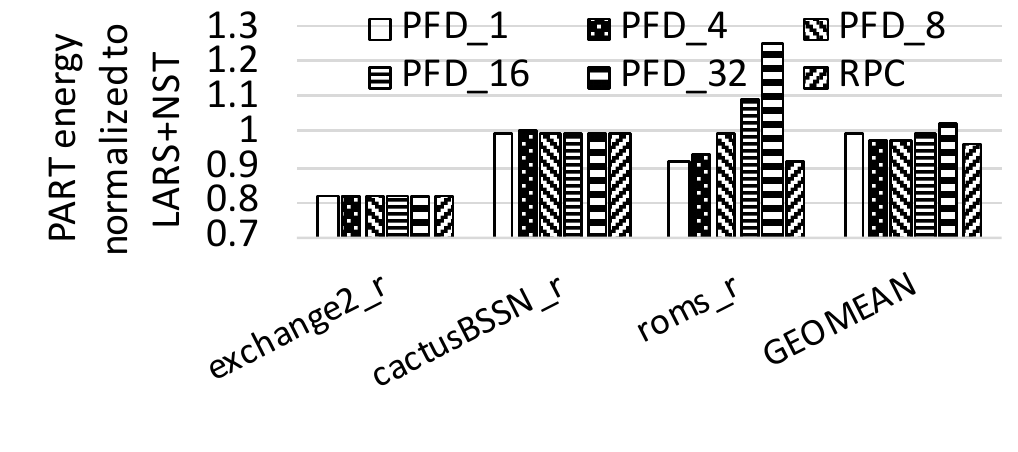}
		\vspace{-10pt}
		\caption{Energy}
		\label{fig:energy_larsnst}
    \end{subfigure}
    \begin{subfigure}{\linewidth}
		\centering
		\includegraphics[width=0.9\linewidth]{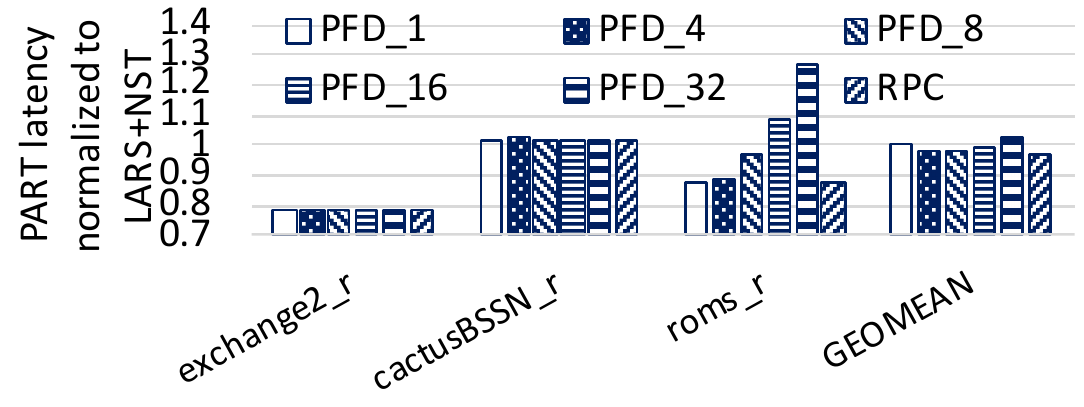}
		\caption{Latency}
		\label{fig:latency_larsnst}
    \end{subfigure}
    \vspace{-5pt}
    \caption{PART with different prefetch scenarios (PFD\_N and RPC) normalized to LARS+NST}   \label{fig:larsnst}
    \vspace{-10pt}
\end{figure}

We further compared PART with LARS+NST to evaluate the improvement when the dynamic prefetch throttling is enabled as in previous work \cite{Heirman18}. For brevity, Figure \ref{fig:larsnst} compares PART to LARS+NST using a subset of notable benchmarks and the geometric mean of all the benchmarks. Figure \ref{fig:energy_larsnst} shows that on average, PART+RPC improved the energy by 3.50\% over LARS+NST, with energy savings of up to 18.77\% for \textit{exchange2}. On the other hand, on average, LARS+NST only improved over LARS+PFD\_16 by 1.43\%. We observed that in STTRAM cache without PART, the dynamic prefetcher (NST) offered minimal energy savings, even if it recovered expired blocks. As shown in Figure \ref{fig:latency_larsnst}, PART+RPC reduced the average latency by 3.59\% compared to LARS+NST, with reductions of up to 21.09\% and 12.23\% for \textit{exchange2} and \textit{roms}, respectively. In the worst case, the latency overhead was 1.60\% for \textit{cactusBSSN}, while the rest of benchmarks benefited from latency reduction. 

In a few cases, PART+RPC did not improve the latency or energy as compared with LARS+PFD\_16 or LARS+NST (for example, for \textit{cactusBSSN}). \textit{CactusBSSN} was one of the benchmarks with a low prefetch percentage in total MSHR requests. As defined in Algorithm \ref{algo:algo1} (line \ref{line:allPF}), the \textit{allPF} in \textit{cactusBSSN} was very low at 0.0002\%. Thus, PART reverts to miss based tuning for \textit{cactusBSSN}, as described in Section \ref{sec:PART}. However, to provide a clear contrast between our work and prior work, we used \textit{expiredPF}-based tuning in all PART+RPC results. For \textit{cactusBSSN}, the RPC table was unable to map the correct prefetch distance for latency or energy improvement. We note, however, that in almost all cases (20 out of 21 benchmarks), PART+RPC outperformed both LARS+PFD\_16 and LARS+NST in both energy and latency. Importantly, we also reiterate that LARS+NST required additional hardware structures to implement the NST prefetcher, whereas RPC's overhead was marginal compared to LARS+PFD\_16, as described in Section \ref{sec:overhead}. The main advantage of PART+RPC is the negligible hardware overhead compared to NST. For instance, NST required seven 32-bit registers for storage \cite{Heirman18}, whereas PART only introduced one additional register to LARS in order to track the number of outgoing MSHR requests, total prefetches, and expired unused prefetches. Overall, PART+RPC reduced the implementation overhead by 54.55\% compared to LARS+NST.

\subsubsection{Exploring the synergy of PART and NST}
We also explored the extent of the benefit, if any, of combining PART with NST (i.e., PART+NST). Figure \ref{fig:synergy} summarizes the energy and latency of PART+RPC and PART+NST normalized to LARS+NST. For brevity, only the geometric mean of all the SPEC CPU 2017 benchmarks are shown. On average, PART+NST improved the energy and latency by 2.75\% and 2.63\%, respectively, compared to LARS+NST, whereas PART+RPC reduced the energy and latency by 3.50\% and 3.59\%, respectively. The results show that while providing dynamic prefetch distance control, NST's increased hardware overhead compared to PART does not translate to energy or latency benefits. In fact, PART still reduced the energy and latency, albeit marginally, while substantially reducing the implementation overheads (Section \ref{sec:part_vs_larsnst}). The results also reveal the promise of a low overhead dynamic prefetch distance control for STTRAM cache based on \textit{expiredPF}. We anticipate that even more energy and latency benefits can be achieved in larger STTRAM caches (such as LLC), and we intend to explore and quantify these benefits in future work.

\subsubsection{Comparison to SRAM with dynamic prefetch distance (SRAM+NST)}

\begin{figure}
\centering
    \begin{minipage}{0.24\textwidth}
	    \centering
		\includegraphics[width=\textwidth]{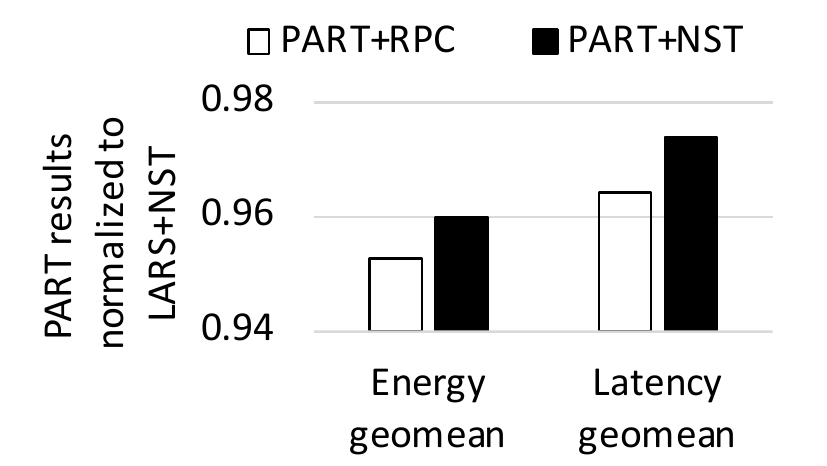}
		\caption{PART normalized to LARS+NST}
		\label{fig:synergy}
    \end{minipage}\hfill
    \begin{minipage}{0.24\textwidth}
		\centering
		\includegraphics[width=\textwidth]{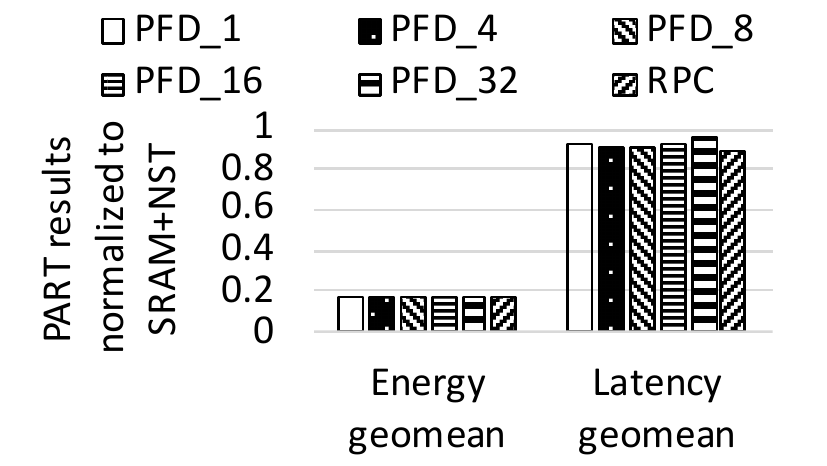}
		\caption{PART normalized to SRAM+NST}
		\label{fig:sram}
    \end{minipage}
\vspace{-10pt}
\end{figure}

We also compare PART to SRAM cache with NST prefetcher enabled (SRAM+NST). Figure \ref{fig:sram} summarizes the energy and latency of PART in the different configurations normalized to SRAM+NST. On average, in all prefetch configurations, PART reduced the energy by more than 80\%. We attribute this reduction largely to the STTRAM's low leakage power (Table \ref{tab:parameters}) and PART's ability to select retention times that satisfied the different applications' cache block requirements. As a result of this specialization, PART was also able to reduce the latency (e.g., by 10.28\% for PART+RPC). As shown in Table \ref{tab:parameters}, STTRAM has advantages in hit latency but not write latency. However, with the help of PART, STTRAM was able to select shorter retention times that satisfy the applications' needs while maintaining write latencies that were close to SRAM. We took a closer look at benchmarks with high write activity, where write requests and miss responses were greater than 40\%, such as \textit{perlbench}, \textit{cactusBSSN}, \textit{povray}, \textit{lbm}, \textit{cam4}, and \textit{fotonik3d}. Our analysis revealed that the synergy of prefetching and PART's retention time selection made the write performance for these benchmarks comparable to SRAM. As a result, the STTRAM cache with PART did not degrade the latency compared to SRAM. 

\section{Conclusions and Future Work}
In this paper, we studied prefetching in reduced retention STTRAM L1 caches. We showed that using \textit{expired\_unused\_prefetches}, and practically, tracking changes in expired prefetches (\textit{expiredPF}) with respect to total prefetches (\textit{allPF}), we could provide an accurate description of the best retention with regards to energy consumption and derive insights into the best prefetch distance. Based on these insights, we proposed prefetch-aware retention time tuning (PART) and retention time based prefetch control (RPC) to predict the best retention time and the best prefetch distance during runtime. Experiments show that PART+RPC can reduce the average cache energy and latency by 22.24\% and 24.59\%, respectively, compared to a base architecture, and by 3.50\% and 3.59\%, respectively, compared to prior work, while reducing the implementation hardware overheads by 54.55\%. For future work, we plan to explore the implications of PART on shared lower level caches and in the presence of workload variations.

\section*{Acknowledgement}
This work was supported in part by the National Science Foundation under grant CNS-1844952. Any opinions, findings, and conclusions or recommendations expressed in this material are those of the authors and do not necessarily reflect the views of the National Science Foundation.
	\balance
	\bibliographystyle{IEEEtran}
	\bibliography{refs}
	
\end{document}